\documentclass[11pt,a4paper]{article}
\usepackage{amsmath}
\usepackage{cite}
\begin{document}
\title{Method of functional integration in the problem of line width of 
parametric X-ray relativistic electron radiation in a Crystall}
\author{\bf N.F.Shul'ga\footnote{shulga@kipt.kharkov.ua} \,
\bf M.Tabrizi\footnote{pgsi@postmaster.co.uk} \\
\addtocounter{footnote}{-1}
\footnotemark[\value{footnote}] 
\small \em Institute for Theoretical Physics, National Scientific Center \\
\small \em "Kharkov Institute of Physics and Technology", 
Kharkov, 61108, Ukraine \\
\addtocounter{footnote}{1}
\footnotemark[\value{footnote}] 
\small \em Department of Physics and Technology, 
Kharkov National University,\\
\small \em Kharkov, 61077, Ukraine}
\maketitle
\begin{abstract}
The coherent and non-coherent scattering effects on "backward" parametric X-ray
radiation by relativistic electrons in a crystal on the basis of the method 
of functional integration is investigated. A comparison of contributions of these
effects to parametric X-ray radiation line width has been considered. It is 
shown that in a number of cases the major contribution to the line width of 
parametric X-ray radiation is made by non-coherent multiple scattering.
\end{abstract}

1. The different coherent and interference radiation effects are possible 
when a relativistic electron moves at a small angle with respect to one of the
crystallographic axes in a crystal \cite{HEEDM,HEEMPCM}. Due to these effects the spectral-angular 
radiation density has sharp maxima with high radiation intensity. One of these 
effects is Parametric X-ray Radiation (PXR) \cite{HEEMPCM,NRSURE,UFN,EPIDM}. This radiation is caused by 
particle field scattering on nonuniformities of lattice permitivity and mainly 
concentrated in direction close to the Bragg direction of particle field reflection
from crystalline planes of atoms. Of special interest is "backward" PXR when an electron
falls at a small incidence angle to one of the crystallographic axes ($z$-axis), beacuse 
in this case the contribution of other kinds of radiation such as coherent and 
channeling radiation are considerably suppressed. The coherent and channeling radiation
are largely concentrated in the line of particle motion. Narrow lines then appear in the spectral-angular 
radiation density as a result of the interference of irradiated waves by electron
at interaction with crystalline planes of atoms oriented normally to $z$-axis.
Experimental studies of PXR line widths have shown \cite{EPIDM} that the line widths were,
however, much larger than the natural PXR line width. Performed analysis of 
experimental data in \cite{EPIDM,JETPL} is shown that the multiple scattering must make considerable
influence on line width of PXR.

In \cite{JETPL} a theory of the line width of "backward" PXR based on the method of functional
integration was suggested. On the basis of this theory in \cite{JETPL} the simplest case of the
multiple scattering effect on the PXR spectral-angular characteristics was considered
in which multiple scattering was studied only in continuous field of the crystal
atomic strings.

This paper is devoted to investigation of non-coherent multiple scattering 
effect of particles on the line width in a crystal. Offered averaging method
in \cite{JETPL} over PXR spectral-angular density is generalized to the case, when coherent 
and non-coherent effects in particle multiple scattering on atomic strings are took 
into account. It is shown that this problem in many respects is similar to the
problem of the Landau-Pomeranchuk-Migdal effect of multiple scattering on coherent
bremsstrahlung of high energy electrons in a crystal and in an amorphous medium
\cite{HEEDM,DAN1,DAN2,SPJETP,SPJETP1}. On the basis of obtained formulas a comparison of contributions of 
coherent and non-coherent multiple scattering of particles in a crystal to PXR line 
width is verified. It is considered the case for which the non-coherent multiple scattering 
makes the major contribution to the PXR line width.

2. We shall investigate "backward" PXR when relativistic electrons fall at a small 
incidence angle $\psi$ with respect to one of the crystallographic axes ($z$-axis). 
Narrow lines then appear in the spectral-angular radiation density caused by
particle field reflection from the crystallographic planes of atoms oriented transversely
to $z$-axis on equal distance $a$ from each other. Spectral-angular radiation density
corresponding to the line width frequency $\omega_{n}$ is given by the equation \cite{JETPL}
\begin{equation}
\frac{dE}{dod\omega}=\frac{e^{2}\omega_{n}^{2}}{4\pi^{2}a^{2}}
\left| \epsilon_{\omega_{n},g} \right|^{2} \left| \int\limits_{0}^L dt \, e^{2i(\omega-\omega_{n})}
\frac{\boldsymbol{\theta}-\boldsymbol{\psi}-\frac{\partial}{\partial \boldsymbol{\mu}}}
{\gamma^{-2}+(\boldsymbol{\theta}-\boldsymbol{\psi})^{2}-
\frac{2i}{\omega_{n}}
\frac{\partial}{\partial t}}\Phi(\mathbf{v}_{\perp}(t)) 
\right|_{\mu\rightarrow 0}^{2}
\end{equation}
where $L$ is target thickness, $\boldsymbol{\psi}=(\psi_{x},0)$ is two-dimensional vector
in the $(x,y)$ plane normally to $z$-axis defining angle $\psi=|\boldsymbol{\psi}|$
between falling particle velocity vector and $z$-axis  (it is proposed that beam 
falls to the crystal in the $(x,y)$ plane) and $\boldsymbol{\theta}=(\theta_{x},0)$
is the angle at which radiation occurs.(we are interested in radiation in the $(x,y)$
plane in the region of angles $\theta_{x}$ close to the direction of Bragg 
reflection of waves; $\boldsymbol{\theta}$ counts off from perpendicular to the 
crystal surface on which beam falls).  

The value $\epsilon_{\omega_{n},g}$ in (1) is the Fourier component of lattice permittivity. 
In our case the permittivity nonuniformity along $x$- and $y$-axes is inessential
to radiation. $\epsilon_{\omega_{n},g}$ is, therefore, formed only by permittivity
non-uniformity along the $z$-axis
\begin{equation}
\epsilon_{\omega_{n},g}=\int_{0}^a dz \, \epsilon_{\omega_{n}}\exp(-igz) \quad ,
\end{equation} 
where $g=\frac{2\pi n}{a}$ and $n$ are integers.

The functional $\Phi(\mathbf{v}_{\perp}(t))$ in (1) determines influence of
multiple scattering on PXR spectral-angular density
\begin{equation}
\Phi(\mathbf{v}_{\perp}(t))=\exp \left\{ \boldsymbol{\mu}\mathbf{v}_{\perp}(t)+
\frac{i\omega_{n}}{2}\int_{0}^t dt^\prime \mathbf{v}_{\perp}^2 (t^\prime)-i\omega_{n}
(\boldsymbol{\theta}-\boldsymbol{\psi})\int_{0}^t dt^\prime \mathbf{v}_{\perp}(t^\prime) 
\right \},
\end{equation}
where $\mathbf{v}_{\perp}(t)$ is the transverse component of electron velocity
vector at time $t$. 
The value $\omega_{n}$ in our geometry is determined by relation
\begin{equation} 
\omega_{n}=vg\cos\psi(1+v\cos(\theta+\psi))^{-1}.
\end{equation}

Eq. (1) was derived taking into account both target thickness and small deviations
of the particle trajectory in crystall from linear one. Obtaining (1) it was
supposed that the medium permittivity is close to unity and also photon absorption
was neglected. The finite size of crystal leads us, as is well \cite{EPIDM,JETPL}, to the natural
line width of PXR. Small deviations of electron trajectory in crystal from linear one 
is caused by both particle plane channeling effect and multiple scattering on 
non-uniformities of lattice potential. The plane channeling effect is absent when
electrons fall at a small incidence angle with respect to one of the crystallographic
axes (see \cite{HEEDM} for example). In this case multiple scattering on atomic strings of crystal parallel to the $z$-axis
leads us to the deviation of particle trajectory from linear one.

Under conditions $|\boldsymbol{\theta}-\boldsymbol{\psi}|\gg|\mathbf{v}_{\perp}(t)|$
the dependence of preexponential factor on random value $\mathbf{v}_{\perp}(t)$
can be ignored. Such dependence in (1) appears as a result of action of differential
operators on the functional $\Phi(\mathbf{v}_{\perp}(t)$. In this simplest situation
formula (1) takes the next form
\begin{equation}
\frac{dE}{dod\omega}=\frac{e^{2}\omega_{n}^{2} \left| \epsilon_{\omega_{n},g} \right|^{2}}
{4\pi^{2} a^{2}}\frac{(\theta-\psi)^{2}}{\left[ \gamma^{-2}+(\theta-\psi)^{2} \right]^{2}}
\left| \int_{0}^{L}dt \, e^{2i(\omega-\omega_{n})t}\Phi(\mathbf{v}_{\perp}(t)) \right|^{2}
\end{equation}

Eq. (5) should be averaged over random angle value of particle scattering
in crystal. Separating out in (5) function subject to averaging we obtain the 
following formula for average value of PXR spectral-angular radiation distribution
\begin{equation}
\langle\frac{dE}{dod\omega}\rangle=\frac{e^{2}\omega_{n}^{2} \left| \epsilon_{\omega_{n},g} \right|^{2}}
{4\pi^{2} a^{2}}
\frac{(\theta-\psi)^{2}}{\left[ \gamma^{-2}+(\theta-\psi)^{2} \right]^{2}} L^{2}
F(L,\omega-\omega_{n}) \quad,
\end{equation}
where 
\begin{equation}
F(L,\omega-\omega_{n})=\frac{2}{L^{2}}\mathrm{Re}\int_{0}^Ldt\int_{0}^tdt^{\prime} \, e^{2i(\omega-
\omega_{n})(t-t^{\prime})}\langle\Phi(t,t^{\prime})\rangle
\end{equation}
and
\begin{equation}
\Phi(t,t^{\prime})=\exp\left\{\frac{i\omega_{n}}{2}\int_{t^{\prime}}^td\tau\,
\mathbf{v}_{\perp}^{2}(\tau)-i\omega_{n}(\boldsymbol{\theta}-\boldsymbol{\psi})
\int_{t^{\prime}}^td\tau\mathbf{v}_{\perp}(\tau)\right\}
\end{equation}

Relativistic electron scattering in a crystal when an 
electron falls at a small incidence angle $\psi$ to crystallographic axis 
is caused by both coherent and non-coherent particles scattering on lattice
atoms. Coherent scattering is mainly along azimuthal angle in the $(x,y)$ plane,
orthogonal to the $z$-axis \cite{HEEDM}. A redistribution of particles over this angle occurs
as a result of multiple scattering by different atomic strings. If $\psi\gg\psi_{c}$, 
the multiple scattering is a Gaussian process with the mean square
of multiple scattering angle $\overline{\theta^2}=q_{c}L$ (here $\psi_{c}=\sqrt{\frac{4Ze^{2}}{\epsilon d}}$ 
is the critical angle of axial channeling, $d$ is the interatomic distance along $z$-axis, $\epsilon$ is
the particle energy) \cite{HEEDM}. Non-coherent scattering at $\psi\gg\psi_{c}$ mainly occurs as a result 
of particle scattering angle by thermal displacement of atom positions in lattice. 
In this case the mean square of multiple scattering angle is close to the value of this one 
in an amorphous medium 
$\overline{\theta_a^2}=q_{a}L$. Thus, particle trajectory deviation along the $y$-axis,
orthoganal to the $(z,\mathbf{v})$ plane, is caused by both coherent and 
non-coherent particle scattering in crystal with mean square of multiple scattering
angle $\overline{\theta_y^2}=q_{y}L$ where $q_{y}=q_{c}+q_{a}/2$. Particle deviation
then along $x$-axis is caused by non-coherent scattering with mean square of multiple 
scattering angle $\overline{\theta_x^2}=q_{x}L$, where $q_{x}=q_{a}$. 

Random values $v_{\perp x}(\tau)$ and $v_{\perp y}(\tau)$ in (8) can be factorized.
The average value of function $\Phi(t,t^{\prime})$ then can be written as the 
prudoct of average magnitudes $\langle\Phi(t,t^{\prime})\rangle=
\langle\Phi_{x}\rangle \cdot \langle\Phi_{y}\rangle$. Since in our case scattering
process is a Gaussian, we can write $\langle\Phi_{x}\rangle$ in terms of functional
integration
\begin{equation}
\begin{split}
\langle\Phi_{x}\rangle=\lim_{N \rightarrow \infty} \int\cdots\int
\frac{dv_{1}\cdots dv_{N}}{(2\pi q_{x} \Delta)^{N/2}}
\exp \biggl\{ -\frac{v_{1}^2}{2q_{x}\Delta}      
\cdots-\frac{(v_{N}-v_{N-1}^{2})}{2q_{x}\Delta}- \\
\begin{align}
-i\omega_{n}\theta_{r}\Delta\sum_{n=k}^{N}v_{n}
+\frac{i\omega_{n}}{2}\Delta\sum_{n=k}^N v_{n}^2 
\biggr\}
\end{align}
\end{split}
\end{equation} 
where $t=N\Delta$, $t^{\prime}=k\Delta$, $v_{n}$ is scattering angle at time 
$t_{n}=n\Delta$ and $\theta_{r}=\theta-\psi$. (Here we use $v\approx 1$.)

In our geometry $\theta$ and $\psi$ are situated in the $(x,y)$ plane. In this 
case the value $\langle\Phi_{y}\rangle$ can be derived from $\langle\Phi_{x}\rangle$
if we replace $q_{x}$ in $\langle\Phi_{x}\rangle$ by $q_{y}$ and use $\theta_{r}=0$.

The functional integral (9) has the same structure as the corresponding integral
in the theory of Landau-Pomeranchuk-Migdal multiple scattering effect on bremsstrahlung
of high energy electrons in an amorphous medium. Integral (9) can be therefore
calculated by the method developed in \cite{SPJETP,SPJETP1}. We then obtain function 
$F(L,\omega-\omega_{n})$ in the next form
\begin{equation}
\addtocounter{equation}{1}
\begin{split}
F(L,\omega-\omega_{n})=2 \mathrm{Re}\int_{0}^1 dx\int_{0}^x du \, e^{2i(\omega-
\omega_{n})Lu} \; \times \\
\times \; \frac{1}{\sqrt{B_{x}B_{y}}}\mathrm{exp}\left\{-\alpha^{2}\sigma_{x}^{2}
\frac{u^{2}(x-2u/3)}{B_{x}}\right\}
\end{split}
\end{equation}
where $\sigma_{x}=\omega_{n}q_{x}L^{2}$, $\alpha^{2}=\frac{\theta_{r}^{2}}
{q_{x}L}$, $\sigma_{y}=\omega_{n}q_{y}L^{2}$, $B_{y}=1-i\sigma_{y}u(x-u)+ \\
+\frac{\sigma_{y}^{2}}{3}u^{3}(x-u)$ 
and $B_{x}=1-i\sigma_{x}u(x-u)+\frac{\sigma_{x}^2}{3}u^{3}(x-u)$.\\
Here we put in dimensionless quantities $u=(t-t^{\prime})/L$ and $x=t/L$.

3.Eq. (10) shows that the influence of multiple scattering on "backward" PXR
is determined by the parameters $\sigma_{x}$, $\sigma_{y}$ and $\alpha$. We shall consider
some extreme cases.

If $\sigma_{x} \rightarrow 0$ and $\sigma_{y} \rightarrow 0$ the multiple scattering
influence on PXR can be ignored. The function $F=F_{0}$ then determines the natural
line width of PXR
\begin{equation}
F_{0}=\frac{\sin^{2}(\omega-\omega_{n})L}{\left((\omega-\omega_{n})L\right)^{2}}
\end{equation}
The line width in this case is given by $\Delta\omega\sim 1/L$ to within the order
of magnitude.

At $\sigma_{x}\rightarrow 0$ Eq. (10) gets over in corresponding results 
\cite{JETPL}.(we note that $q_{y}$ is related in the Eq. (13) of the work 
\cite{JETPL} by means of ratio $q_{y}=q/2$. The value $\overline{\theta^{2}}$ then
in the \cite{JETPL} must be determined by the ratio $\overline{\theta^{2}}=qL/2=q_{y}L$) 
In this case if $\sigma_{y}\gg1$ the line width to within the order
of magnitude is given by  
\begin{equation}
\Delta\omega\sim\frac{1}{L}\left(\frac
{\sigma_{y}^{2}}{3}\right)^{1/3}
\end{equation}
If $\sigma_{y}\sim\sigma_{x}\gg1$ the major contribution to integral over $u$
in (10) at $\alpha\gg1$ can be made by 
$u_{eff}\sim1/\alpha\sigma_{x}\ll1$.(We use here that the inequality $\alpha\gg1$ is the validity condition
of Eq. (5).) Neglecting the values of order $1/\alpha\sigma_{x}$
we can rewrite (10) in the next form
\begin{equation}
F(L,\omega-\omega_{n})=2\mathrm{Re}\int_{0}^{1}dx\int_{0}^{x}du\, e^{2i(\omega-
\omega_{n})Lu}\exp{\left(-\alpha^{2}\sigma_{x}^{2}u^{2}x\right)}.
\end{equation}

According to (13) the line width to within the order of magnitude is given by
\begin{equation}
\Delta\omega\sim\frac{1}{L}\alpha\sigma_{x}.
\end{equation}
Thus, at $\sigma_{x}\sim\sigma_{y}\gg1$ and $\alpha\gg1$ the non-coherent multiple 
scattering causes substantial broadening of PXR lines compared with natural line
width $\Delta\omega\sim1/L$. Under the conditions of experiment \cite{EPIDM} in which electrons
with the energy $\epsilon=855$ MeV moved in a silicon crystal at a small angle to 
the $\left<111\right>$ axis, the value $\alpha\sigma_{x}$ at $\theta_{r}\sim5\cdot
10^{-3}$ rad is $\alpha\sigma_{x}\sim40n$. 
\newpage 
\noindent \large \bf Acknowledgement  \\
\normalsize \rm N.F.Sh. thanks Professor H.Backe from the Mainz University
for discussion of measurements and interpretation of "backward" line width. M.T.
thanks the Nuclear Energy Society of France (SFEN) for financial support.

\end{document}